\pdfoutput=1

\documentclass[twocolumn,showpacs,amssymb,aps,nofootinbib,floatfix,superscriptaddress]{revtex4-1}

\usepackage{epsfig}

\begin{document} 
\hbadness=10000

\title{Signatures of alpha clustering in light nuclei from relativistic nuclear collisions}

\author{Wojciech Broniowski}
\email{Wojciech.Broniowski@ifj.edu.pl}
\affiliation{Institute of Physics, Jan Kochanowski University, 25-406 Kielce, Poland}
\affiliation{H. Niewodnicza\'nski Institute of Nuclear Physics PAN, 31-342 Cracow, Poland}

\author{Enrique Ruiz Arriola}
\email{earriola@ugr.es}
\affiliation{Departamento de F\'{\i}sica At\'{o}mica, Molecular y Nuclear and Instituto Carlos I de  F{\'\i}sica Te\'orica y Computacional, 
 Universidad de Granada, E-18071 Granada, Spain}

\date{ver. 2, 9 February 2013}  

\begin{abstract}
We argue that relativistic nuclear collisions may provide experimental
evidence of $\alpha$ clustering in light nuclei.  A light $\alpha$-clustered
nucleus has a large intrinsic deformation. When collided against a
heavy nucleus at very high energies, this deformation transforms into
the deformation of the fireball in the transverse plane. The
subsequent collective evolution of the fireball leads to harmonic flow
reflecting the deformation of the initial shape, which can be measured
with standard methods of relativistic heavy-ion collisions. We
illustrate the feasibility of the idea by modeling the
${}^{12}$C--${}^{208}$Pb collisions and
point out that very significant quantitative and qualitative
differences between the $\alpha$-clustered and uniform ${}^{12}$C nucleus
occur in such quantities as the triangular flow, its event-by-event
fluctuations, or the correlations of the elliptic and triangular
flows.  The proposal offers a possibility of studying low-energy
nuclear structure phenomena with ``snapshots'' made with relativistic heavy-ion collisions.
\end{abstract}


\maketitle
%%%%%%%%%%%%%%%%%%%%%%%%%%%%%%%

In this Letter we show that the nuclear collisions in the ultrarelativistic domain may reveal, via harmonic flow, 
geometric $\alpha$ clustering structure of light nuclei in their ground state.
As a particular example we present a
study of in ${}^{12}$C, where a triangular structure
induces a corresponding pattern in the collective flow.

The $\alpha$ cluster model was proposed even before the discovery of
the neutron by Gamow~\cite{gamow1931constitution} and rests on the
compactness, tight binding $B_\alpha/4 \sim 7 {\rm MeV}$, and
stability of the $^4$He
nucleus~\cite{PhysRev.52.1083,PhysRev.54.681,wefelmeier1937geometrisches},
which fits into the SU(4) Wigner symmetry of the quartet
($p \uparrow$, $p \downarrow$, $n \uparrow$, $n \downarrow$) (see,
e.g.,~\cite{blatt19521952theoretical} for an early review
and~\cite{brink2008history} for a historic account, while many
references can be traced
from~\cite{brink1965alpha,freer2007clustered,ikeda2010clusters,beck2012clusters}).
The remaining weak binding per bonding between the $\alpha$-particles,
$V_{\alpha \alpha}/{\rm bond} \sim 2 {\rm MeV}$, accounts for nuclear
binding and makes a molecular picture of light nuclei quite
natural. This suggests a vivid geometric view of the self-conjugate
$A=4n$ nuclei classified by point group
symmetries~\cite{Brink1970143}. For instance, in ${}^{9}$Be the two
$\alpha$ clusters are separated by as much as $\sim$2~fm, ${}^{12}$C
exhibits a triangular arrangement of the three $\alpha$'s $\sim 3 {\rm
fm}$ apart, ${}^{16}$O forms a tetrahedron, etc. 
%This quartetting
%makes nuclear matter at low densities cavitate into
%$\alpha$-particles, a feature displayed in the nuclear surface. 
The condensation of $\alpha$ clusters was described in
\cite{Tohsaki:2001an}  for $^{12}$C and $^{16}$O.  
Clustering of $^{20}$Ne has also recently been described within the
density functional theory~\cite{Ebran:2012uv}. 
Model calculations are verified by comparing to the experimental
binding energies, the elastic electromagnetic form
factor, and the excitation spectra. Experimental evidence for 
clustering comes from fragmentation studies, see, e.g.,~\cite{Zarubin}.

Our basic observation
and the following methodology stems from the fact that the {\em intrinsic}
wave functions of light $\alpha$-clustered nuclei are
deformed~\cite{buendia2001projected}, exhibiting spatial
correlation between the location of clusters.
%
%To illustrate our points, 
Imagine we collide a light $\alpha$-clustered
nucleus against a heavy nucleus at extremely high energies, as in
relativistic colliders (RHIC, LHC) or fixed-target experiments
(SPS). During the almost instantaneous passage of the light nucleus
through the medium, its wave function collapses, revealing the spatial
correlation structure. Let us consider ${}^{12}$C as an example of a triangular $\alpha$-cluster arrangement, colliding with
${}^{208}$Pb.  In a typical collision event, the shape of the created
{\em fireball} in the transverse plane reflects the shape of
${}^{12}$C, washed out to some degree by
different orientations of ${}^{12}$C and 
statistical fluctuations.
% of the locations from the individual NN collisions. 
%
Next, this asymmetric fireball evolves. As the setup is very similar to
that in relativistic heavy-ion collisions, we expect collective
dynamics to properly model the evolution of the compact
dense system, which can be achieved with hydrodynamics (for recent
reviews see~\cite{Heinz:2013th,Gale:2013da} and references therein) or
transport~\cite{Lin:2004en} approaches.  It has been
well established in the studies of relativistic heavy-ion collisions
(and even in d-Au and p-Pb
collisions~\cite{Bozek:2011if,Adare:2013piz,Sickles:2013mua,Bozek:2011if,Bozek:2013ska,Bzdak:2013zma,Qin:2013bha})
that collective dynamics leads to an event-by-event transmutation of
the initial anisotropies, described in terms of the harmonic
transverse-shape coefficients $\epsilon_n$, into the harmonic flow
coefficients in the transverse momentum distributions of the produced
particles, $v_n$. Therefore, applying the well-developed 
methods~\cite{Ollitrault:1992bk,Borghini:2001vi,Voloshin:2002wa,Voloshin:2008dg} of
the harmonic flow analysis successful in relativistic heavy-ion
collisions we may indirectly measure, or assess, the spatial
deformation of the initial state.  In this Letter we argue that the
effects of the $\alpha$ clustering in light nuclei lead to flow effects
which are strong enough to be detectable via ultra-relativistic
nuclear collisions.

The mentioned 
%The density of sources in the transverse plane in each event may be
%decomposed in the Fourier series, $f(\vec{x})=\sum_n \exp( in \phi)
%f_n(\rho)$, with $\phi$ and $\rho$ denoting the polar coordinates.
%One defines the 
eccentricity parameters $\epsilon_n$ and the angles of
the principal axes $\Phi_n$ for a distribution of points in the transverse plane are defined as
%\begin{eqnarray}
$\epsilon_n e^{i n \Phi_n} = \sum_j \rho_j^n e^{i n \phi_j}/\sum_j \rho_j^n$,
%\end{eqnarray}
where $n$ is the rank, $j$ labels the points,
while $\rho_j$ and $\phi_j$ are their polar coordinates.  The measures $\epsilon_2$ and $\epsilon_3$
are referred to as the {\em ellipticity} and {\em triangularity}.  

\begin{figure}[tb]
\centering
\vspace{-2mm}
\includegraphics[angle=0,width=0.34 \textwidth]{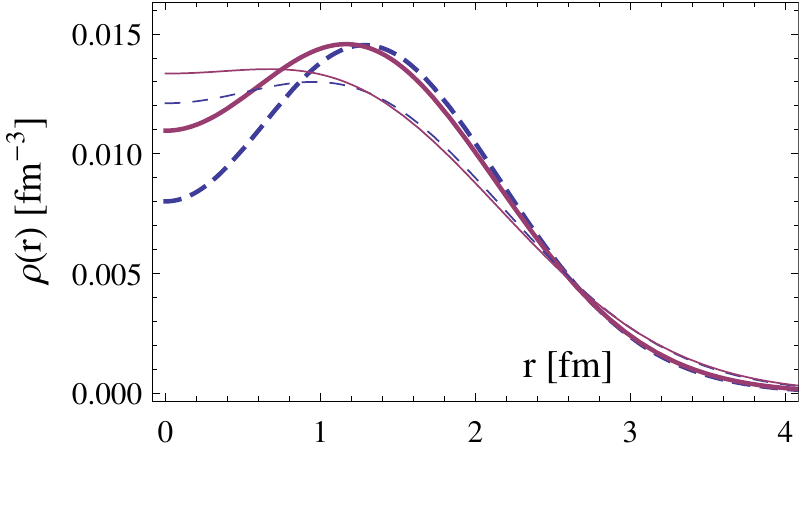}
\vspace{-8mm}
\caption{(Color online) \label{fig:dense} The electric charge density (thin lines) and the corresponding distribution of the centers of 
nucleons (thick lines) in ${}^{12}$C for the data and BEC calculations (dashed lines), and for the FMD calculations (solid lines), plotted against the radius.}
\end{figure}

We focus our present study on ${}^{12}$C (the general program is outlined in conclusions), as it leads
to interesting properties due to large triangularity. We apply
GLISSANDO~\cite{Broniowski:2007nz,*Rybczynski:2013yba} to carry out the Glauber Monte
Carlo ultra-relativistic collisions with ${}^{208}$Pb. The first task
is to properly model the nucleon density of ${}^{12}$C, including the
cluster correlations.  The analysis of data for the elastic
electromagnetic form factor~\cite{DeJager:1987qc} imposes an important
constraint on the charge density~\cite{Chernykh:2007zz}, which leads
to the function indicated with the thin dashed line in
Fig.~\ref{fig:dense} (cf.~Fig.~1 of ~\cite{Chernykh:2007zz}). This
distribution is reproduced with the Bose-Einstein condensation (BEC)
wave function~\cite{Funaki:2006gt}. 
%We use it as a reference 
%and label as {\em case I}. 
%for our calculations presented in this Letter.
On the other hand, calculations based on
fermionic molecular dynamics (FMD)~\cite{Chernykh:2007zz}, which
properly reproduce the binding energy, give a somewhat weaker
clustering, with the density drawn in Fig.~\ref{fig:dense} with a
thin solid line. 
%We refer to it as {\em case II} and use in our
%analysis to also explore the case with a weaker clustering.

To carry out the NN collisions, we need the distribution of centers of
nucleons in ${}^{12}$C. For that purpose we unfold from the ${}^{12}$C
elastic charge form factor the proton charge form factor, assumed in
the Gaussian form ${\rm const}\times\exp \left ( - 3/2 \,
r^2/r_p^2 \right )$ with the charge proton radius squared of
$r_p^2=0.77~{\rm fm}^2$. The resulting densities of centers of
nucleons are plotted in Fig.~\ref{fig:dense} with thick lines for
the BEC case (dashed) and the FMD case (solid). Note a large central depletion in
the distributions, originating from the separation of the $\alpha$
clusters arranged in the triangular configuration.

%%%%%%%%%%%%%%%%%%%%%%%%%%%%%%
%%%%%%%%%%%%%%%%%%%%%%%%%

Technically, we proceed as follows: The centers of the clusters are
placed in an equilateral triangle of side length $l$. The nucleons in
clusters have a Gaussian radial distribution of the form ${\rm
const}\times\exp \left (- 3/2 \, r^2/r_\alpha^2 \right )$, from which
we randomly generate positions of the 12 nucleons, 4 in each
cluster. We take into account the short-distance NN repulsion,
precluding the centers of each pair of nucleons to be closer than the
expulsion distance of 0.9~fm~\cite{Broniowski:2010jd}. Finally, the
distribution of 12 nucleons is recentered such that the center of mass
is at the origin. The parameters $l$ and $r_\alpha$
are optimized such that the thick curves in
Fig.~\ref{fig:dense} are accurately reproduced.   
This apparently crude procedure reproduces not only the one-body densities, but also
semi-quantitatively ($\sim10-20\%$) the pair densities determined by the
multicluster models with state-dependent Jastrow
correlations~\cite{Buendia:2004yt}.  For the unclustered
case we generate the 12 nucleons from a uniform radial distribution of
the form $(a +b r^2) \exp(-c^2 r^2)$, also with short-distance
repulsion and recentering. Again, the parameters are adjusted such that the
thick lines in Fig.~\ref{fig:dense} are reproduced. 

The resulting two-dimensional projections of the obtained intrinsic distributions are
displayed in the left panels of Fig.~\ref{fig:all}. For the clustered (BEC) case the projection plane is defined by the centers 
of the three clusters (the {\em cluster} plane). We note prominent
cluster structures for the BEC case (upper left panel). For the uniform distribution (bottom left panel)
there is, by construction, no clustering. The results obtained for the FMD case (not presented here for brevity) are 
qualitatively similar to the BEC case, with somewhat weaker clustering.

%%%%%%%%%%%%%%%%%%%%%%%%%%%%%%
\begin{figure}[tb]
\centering
\vspace{-7mm}
\includegraphics[angle=0,width=0.3 \textwidth]{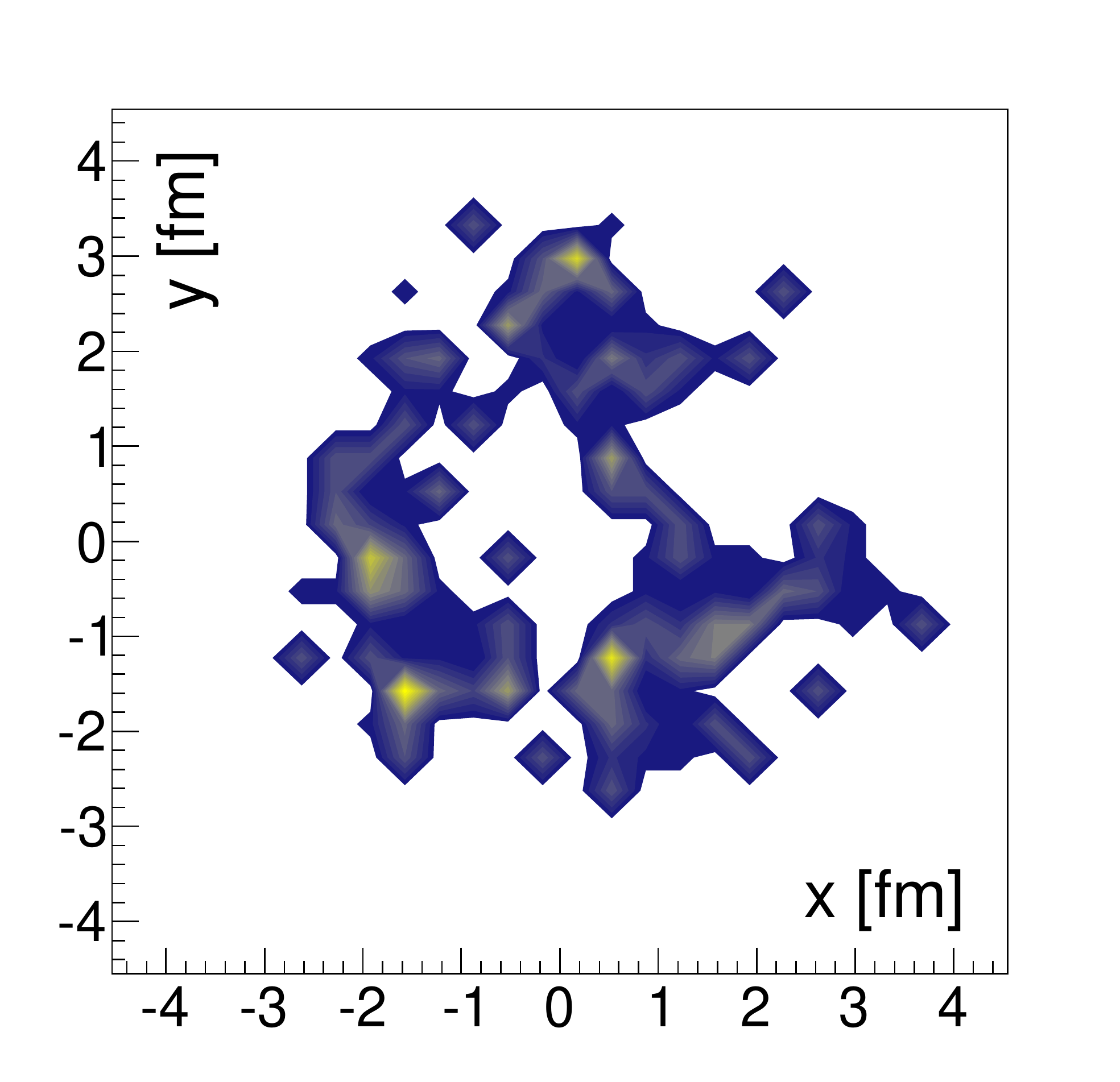}
\vspace{-5mm}
\caption{(Color online) \label{fig:event} Snapshot of a single central ${}^{12}$C--${}^{208}$Pb collision, displaying 
the distribution of sources in the transverse plane, BEC case, $N_{\rm w}=66$, $N_{\rm bin}=93$. In this simulation the transverse and cluster planes were aligned.}
\end{figure}
%%%%%%%%%%%%%%%%%%%%%%%%%

We may use the eccentricities to characterize the intrinsic non-spherical nuclear distributions. 
In the cluster plane
the average triangularity for the ${}^{12}$C distributions equals 0.59 for the BEC and 0.55 for the FMD cases. These are 
substantial, as the extreme value for point-like clusters is 1. The average triangularity
vanishes for the unclustered case. By symmetry, ellipticity is zero.
In the plane perpendicular to the cluster plane the average ellipticity equals 0.61 for the BEC, 0.58 for the FMD, and 0 for the unclustered case,
while the average 
triangularity vanishes by symmetry.

%%%%%%%%%%%%%%%%%%%%%%%%%%%%%%
\begin{figure*}
\centering
\vspace{-7mm}
\hspace{1mm}\includegraphics[angle=0,width=0.3 \textwidth]{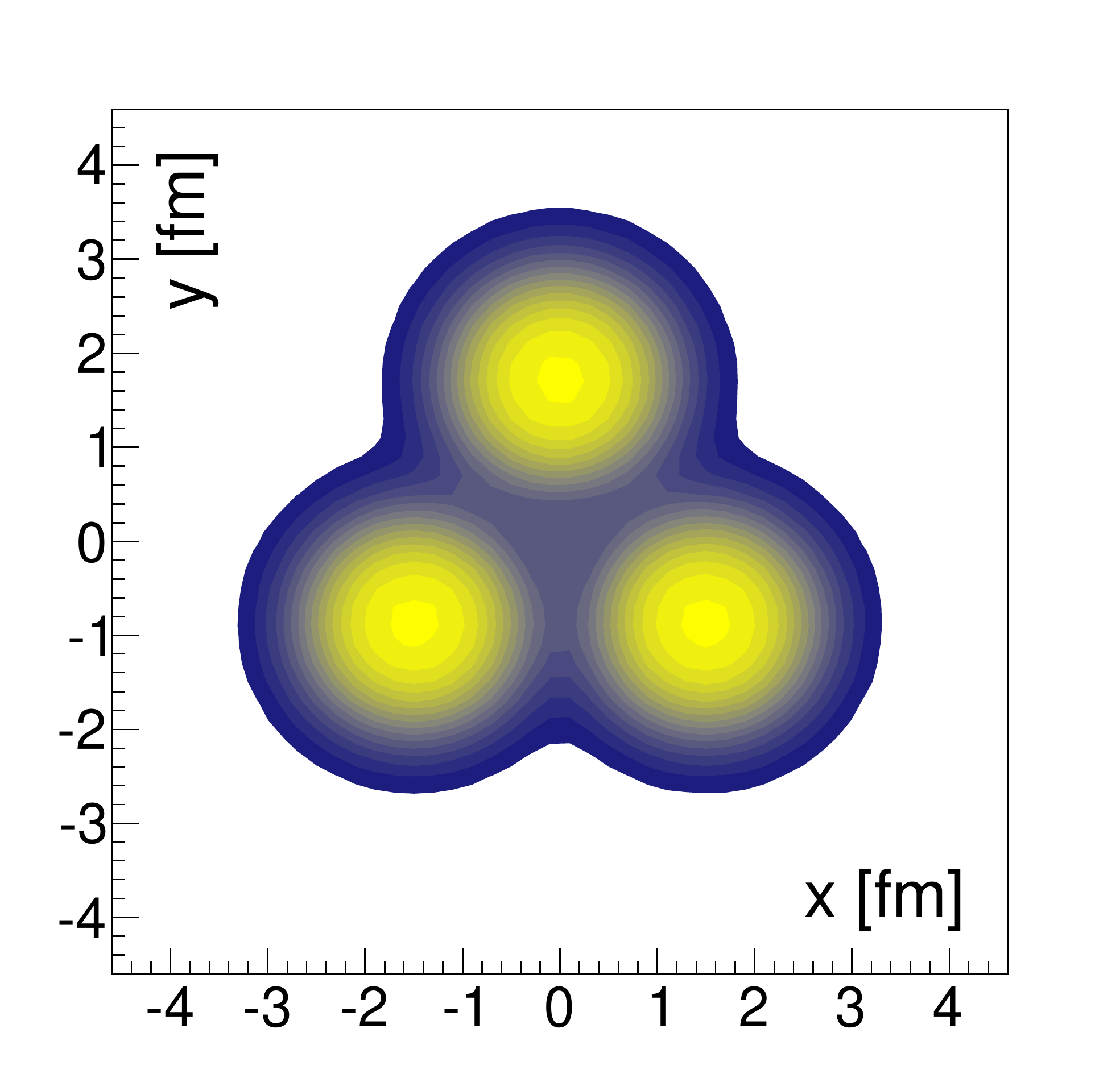}
\includegraphics[angle=0,width=0.3 \textwidth]{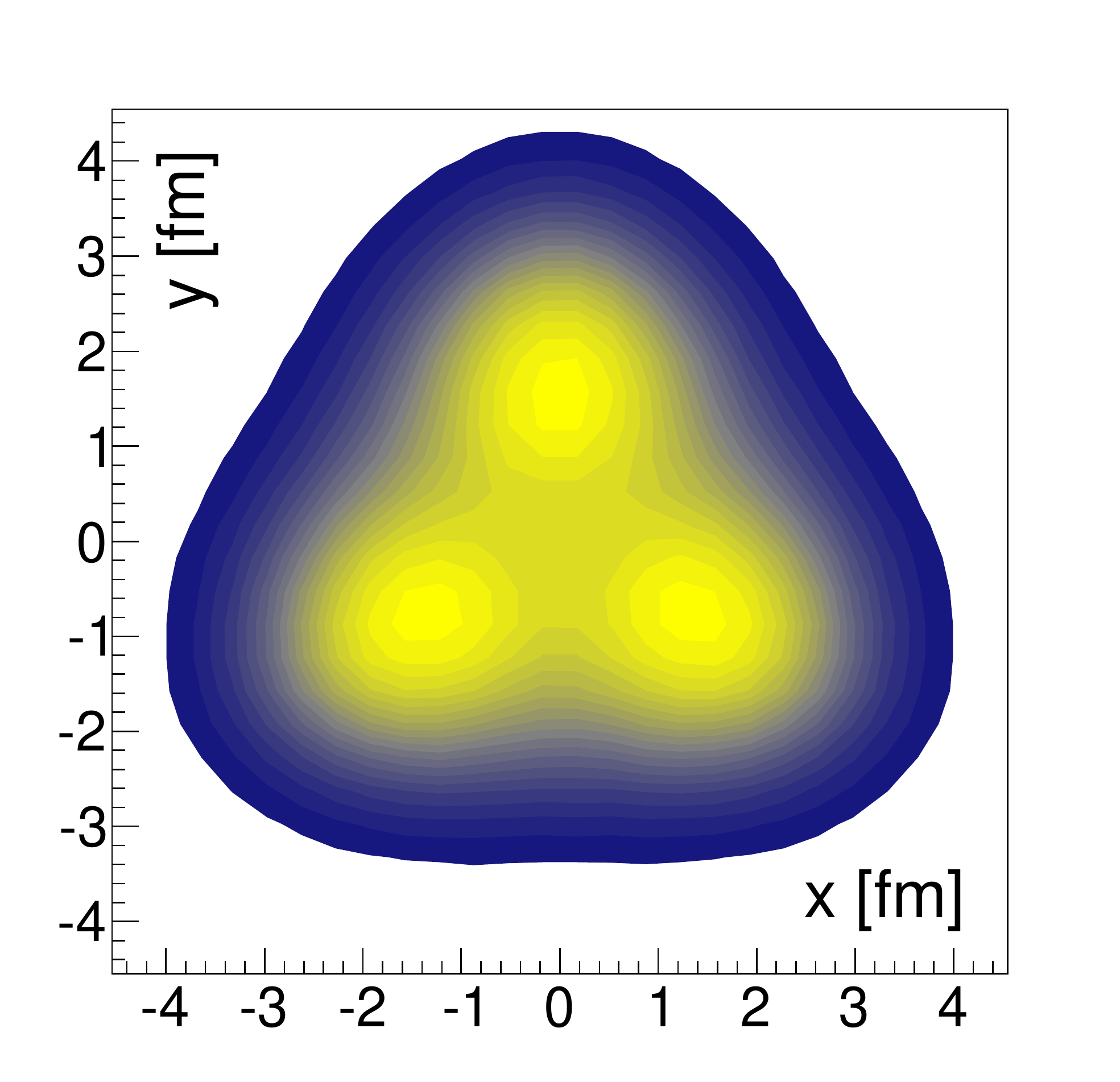}
\includegraphics[angle=0,width=0.3 \textwidth]{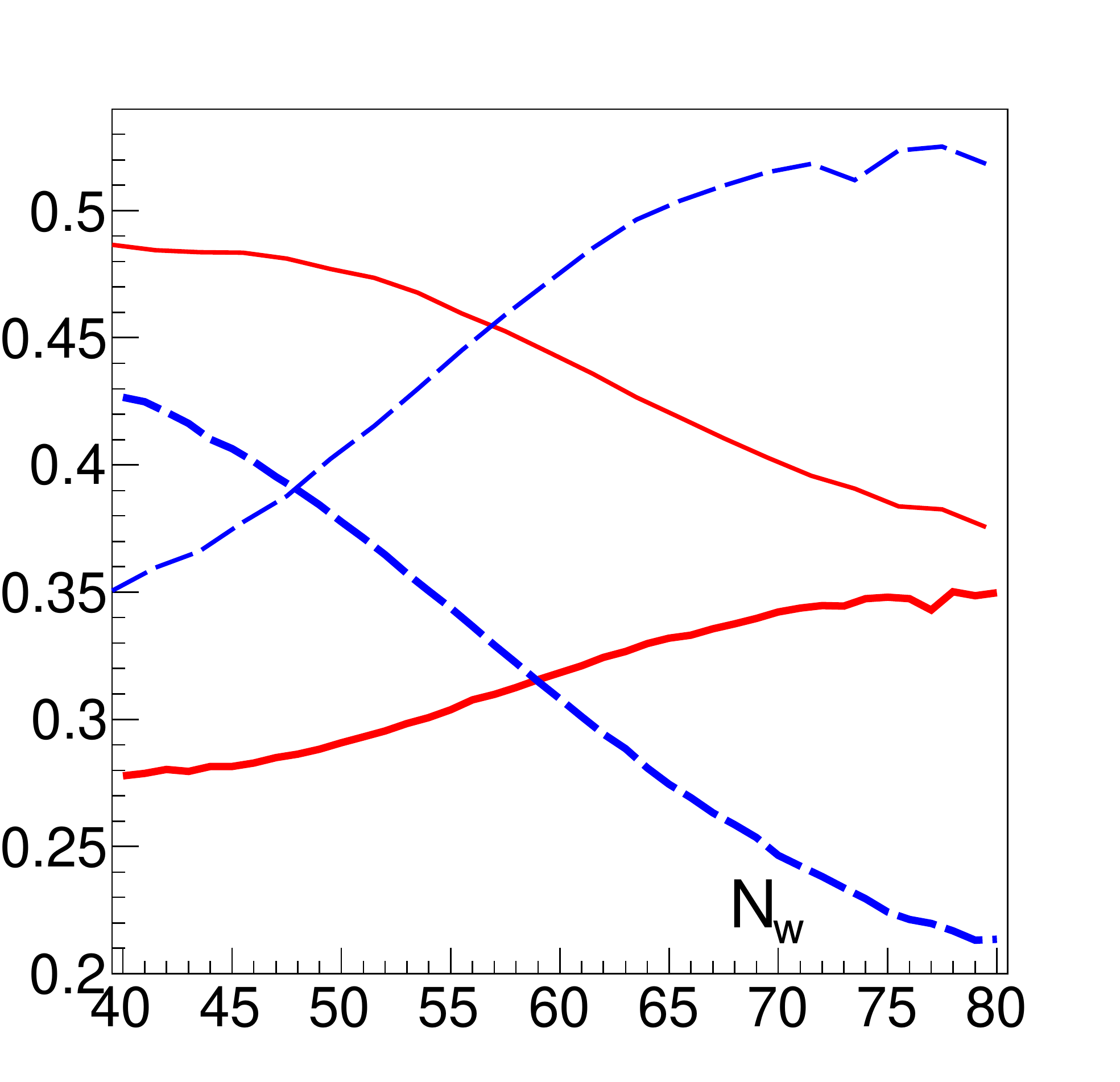} \vspace{-3mm} \\
\includegraphics[angle=0,width=0.3 \textwidth]{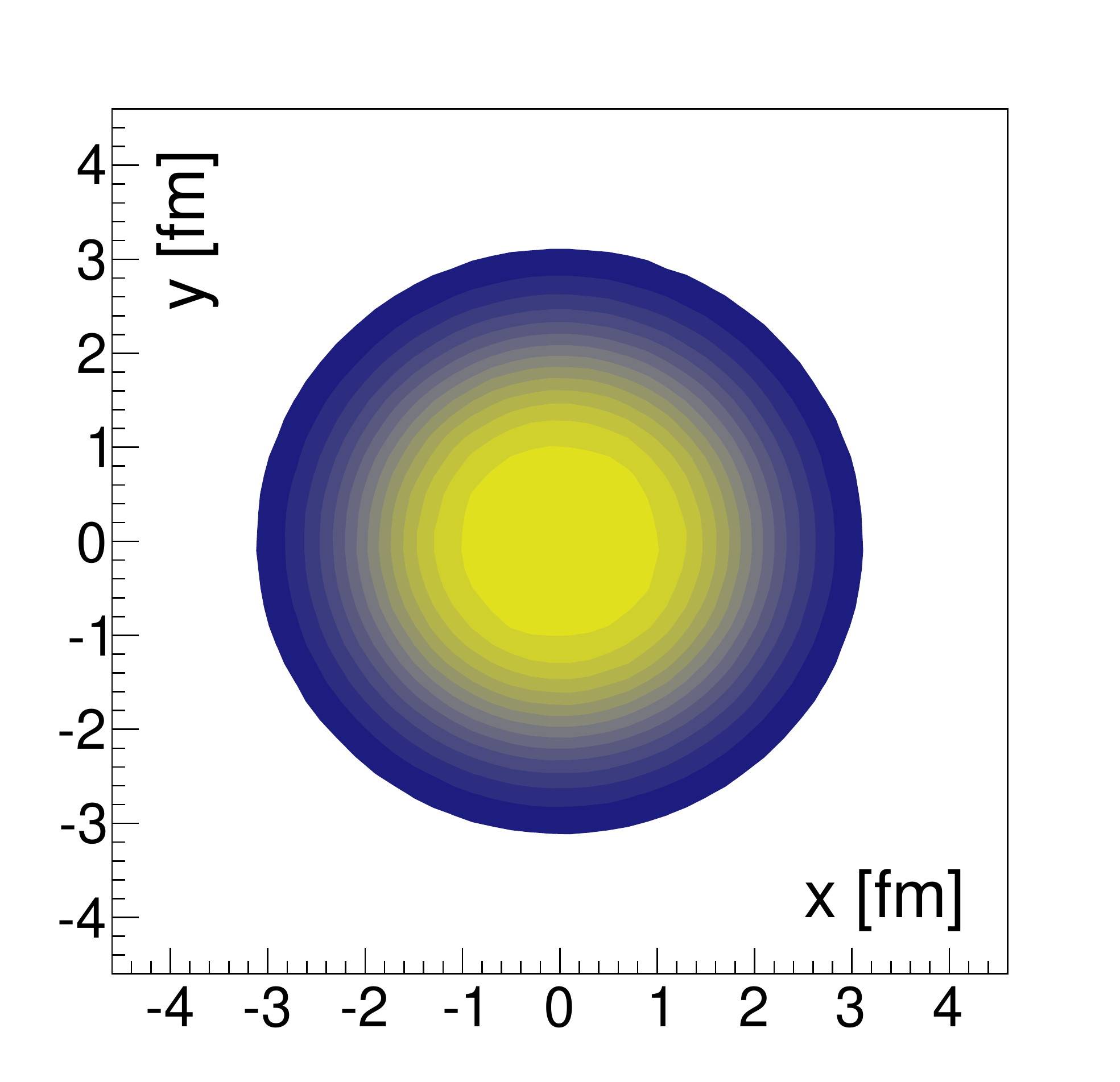}
\includegraphics[angle=0,width=0.3 \textwidth]{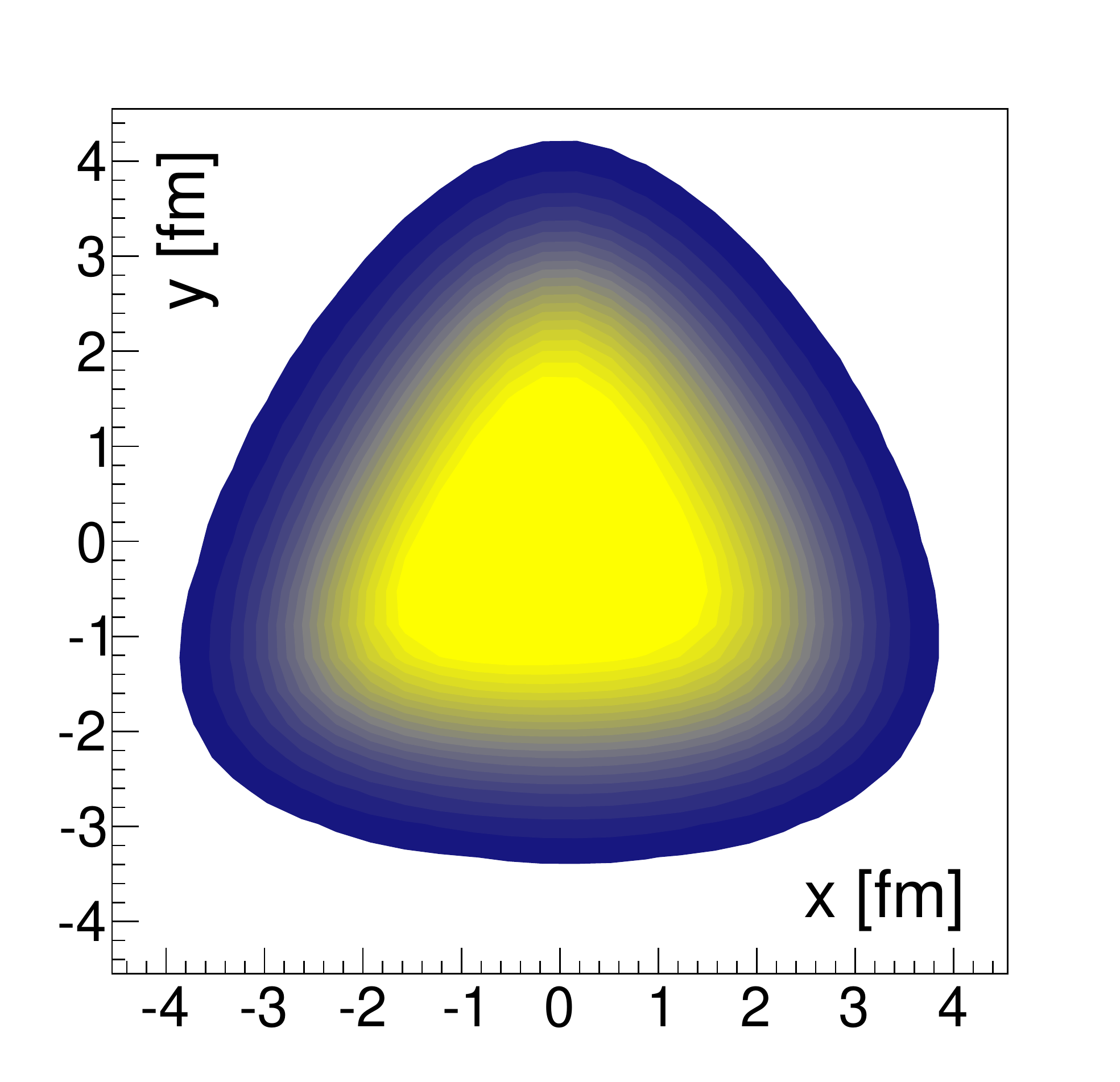}
\includegraphics[angle=0,width=0.3 \textwidth]{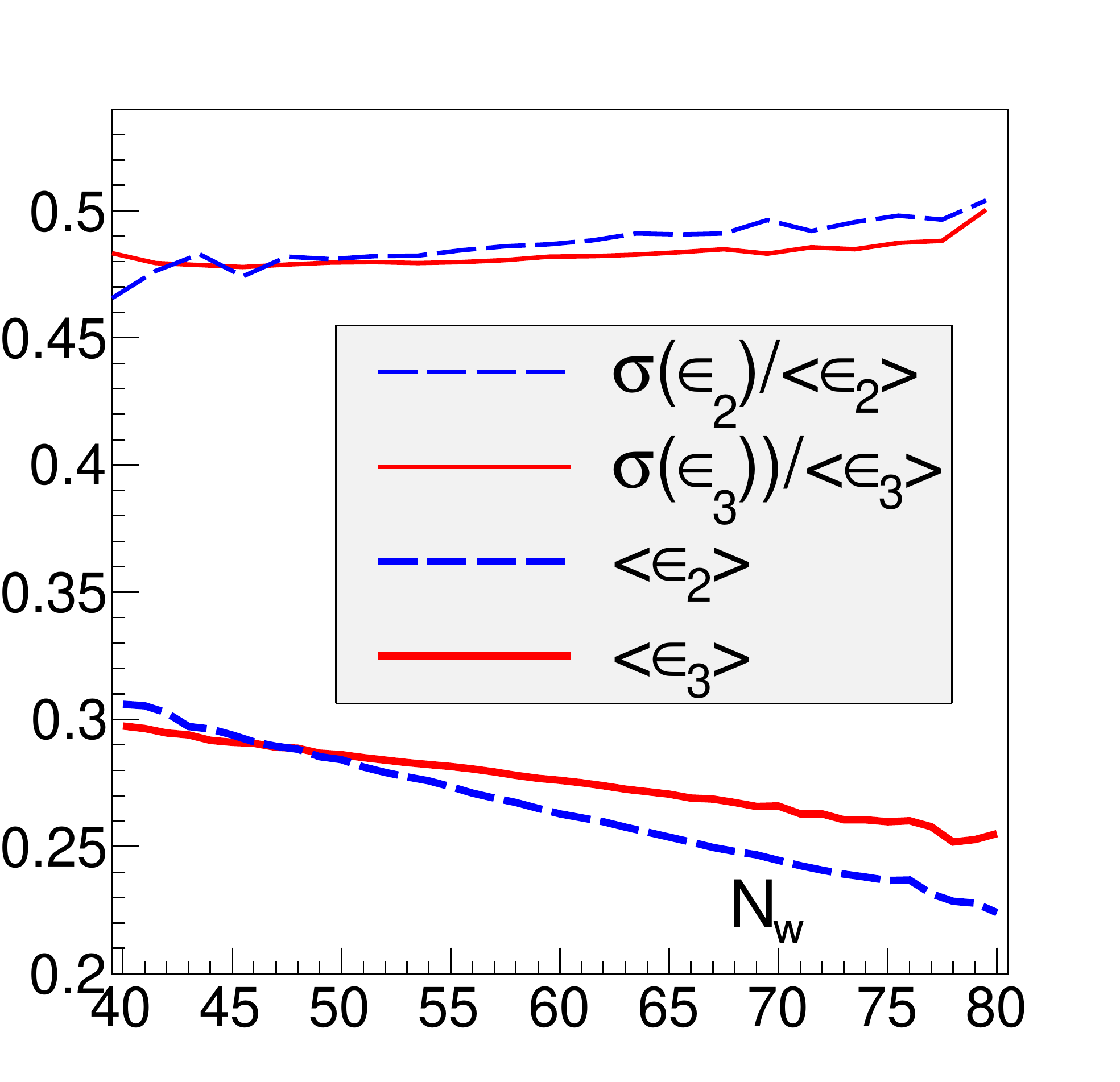} \vspace{-3mm}
\caption{(Color online) \label{fig:all} Glauber Monte Carlo simulations with GLISSANDO for the ${}^{12}$C-${}^{208}$Pb collisions at the SPS 
energy $\sqrt{s_{NN}}=17$~GeV. The top panels correspond to the clustered BEC case, while the bottom panels display the unclustered
case.  The left panels show the intrinsic densities in the ${}^{12}$C
nucleus, the middle panels give the corresponding rank $n=3$ intrinsic
densities of sources in the fireball in the transverse plane for
collisions with a high number of wounded nucleons, $N_w\ge 70$, and the
right panels show the event-by-event statistical properties of the
fireball (average ellipticity, triangularity, and their scaled
standard deviations) as functions of the number of wounded nucleons. See the
text for details.}
\end{figure*}
%%%%%%%%%%%%%%%%%%%%%%%%%

Now we are ready to carry out the collisions with ${}^{208}$Pb, which is prepared in a standard 
way by uniformly generating 208 nucleons from a Woods-Saxon radial 
distribution, with the short-distance repulsion taken into account~\cite{Broniowski:2010jd}. The Glauber mechanism of the reaction
(for a review see, e.g.,~\cite{Florkowski:2010zz}) proceeds through independent high-energy 
collisions of the nucleons from ${}^{12}$C with nucleons from ${}^{208}$Pb. The concepts of wounded nucleons (those that interacted inelastically at 
least once)~\cite{Bialas:1976ed} and the binary collisions turn out to be 
very useful in describing the particle production mechanism. Final multiplicities of the produced 
particles are properly reproduced if the initial parton production is proportional to a combination of the number of wounded nucleons, $N_{\rm w}$, 
and binary collisions,  $N_{\rm bin}$, namely 
$\sim (1-a)/2 N_{\rm w} + a N_{\rm bin}$, which is the {\em mixed} model of Ref.~\cite{Kharzeev:2000ph,Back:2001xy}. 
%The phenomenological parameter $a$ is of the order of 10-15\%. 
For the collision energies corresponding to the top SPS energy of $\sqrt{s_{NN}}=17$~GeV
%, which we take for the 
% results shown in this Letter, the phenomenological parameter 
we take $a=0.12$.
% works well. 
The total NN inelastic cross section is, at this energy,
$\sigma_{NN}^{\rm inel}=32$~mb (all our results do not qualitatively
change when $\sigma_{NN}^{\rm inel}$ is increased up to the LHC values of
$~\sim70$~mb). We use the realistic Gaussian wounding profile in the
simulations~\cite{Broniowski:2007nz}, meaning that the probability
that the two nucleons interact is a Gaussian in their relative impact
parameter, of the width controlled by $\sigma_{NN}^{\rm inel}$. The
wounded nucleons and binary collisions are jointly referred to as {\em
sources}. The outcome of the Monte Carlo simulation is a distribution
of locations of sources in the transverse plane in each event,
$f(\vec{x})=\sum_j \delta(\vec{x}-\vec{x_j})$. In actual applications
the sources are {\em smeared}. This {\em physical} effect is necessary
in preparing the initial condition for hydrodynamics.

A single event of a central (impact parameter equal zero) ${}^{12}$C--${}^{208}$Pb collision is shown in Fig.~\ref{fig:event}. Here we have used the 
clustered ${}^{12}$C BEC distribution and aligned the transverse and the cluster planes (the carbon hits the lead ``flat''). 
The shown collision led to $66$ wounded nucleons and $93$
binary collisions. Note the typical ``warped'' structure following from the stochastic nature of the process, with the underlying three clusters 
structure visible.

The eccentricity coefficients of the fireball have two sources. One comes from the average shape
(for instance, in non-central A-A collisions the overlap almond-shape
region produces $\epsilon_2$, or in the present case the triangular
cluster shape of ${}^{12}$C generates triangularity), but, in
addition, there is a component from fluctuating positions of the
finite number of $N$ sources. This fluctuating
component~\cite{Miller:2003kd,Manly:2005zy,Voloshin:2006gz,Broniowski:2007ft,Alver:2010gr}
is suppressed with $N$.  The {\em intrinsic} density of sources of rank $n$
is defined as the average over events, where the distributions in each
event have aligned principal axes: $f_n^{\rm intr}(\vec{x})=\langle
f(R(-\Phi_n) \vec{x})\rangle$. Here the brackets indicate averaging
over events and $R(-\Phi_n)$ denotes an inverse rotation by the
principal-axis angle in each event.
%
%As mentioned, the distributions of sources fluctuates widely from event to event. If, however, we overlay these densities in a suitable way, we 
%notice the imprints of the ${}^{12}$C distribution. For instance, we may align the angle of the triangular principal axes $\psi_3$ in each event and 
%combine the rotated that way distributions of sources over many events. 
The result of this procedure for generating the intrinsic fireball densities of rank $n=3$
is shown in the middle panels of Fig.~\ref{fig:all} for high-multiplicity  collisions (with more than 70
wounded nucleons). In these simulations the orientation of the ${}^{12}C$ nucleus is in general completely random, however the cut 
$N_{\rm w}>70$ (imposed for better visibility of the clustering effect) selects preferentially the alignment of the transverse and cluster planes (see the following paragraph). 
We note the clear traces of the three $\alpha$ clusters in the middle top panel, while the uniform case (middle bottom panel) is smooth. 
Note, however, that even the uniform case develops some triangularity, which is entirely due to fluctuations~\cite{Alver:2010gr}.

%%%%%%%%%%%%%%%%%%%%%%%%%
\begin{figure}
\centering
\vspace{-7mm}
\includegraphics[angle=0,width=0.3 \textwidth]{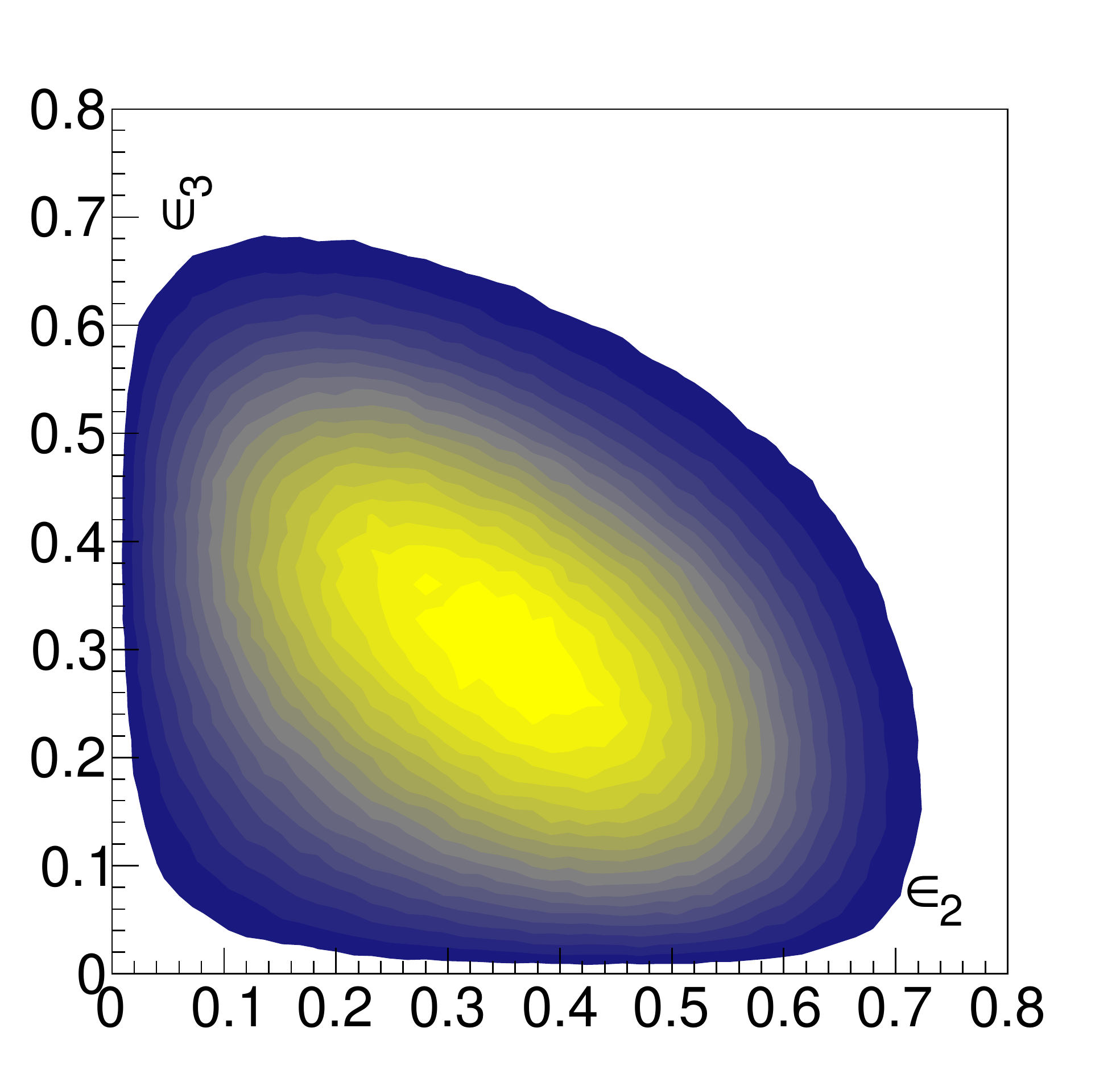} \vspace{-3mm} \\
\caption{(Color online) Correlations between
ellipticity and triangularity for events with $N_w\ge 40$ for the BEC clustered case. \label{fig:corr}}
\end{figure}
%%%%%%%%%%%%%%%%%%%%%%%%

%For the considered case of ${}^{12}$C collisions, 
There is a geometric link 
between the orientation of the clustered intrinsic distribution of ${}^{12}$C and the multiplicity
of the collision. When the transverse plane and the cluster plane are aligned, the clusters hit the Pb nucleus 
side-by-side and create most damage, i.e. produce the largest number of sources. 
In that orientation we have on the average the highest triangularity 
and the lowest ellipticity (which comes only from fluctuations). 
On the other hand, when the cluster plane is perpendicular to the transverse plane, we have the opposite behavior: 
lowest multiplicity, small triangularity,
and large ellipticity, which in this case picks up a contribution from the elongated shape of the fireball in the transverse plane. These 
orientation-multiplicity correlations are crucial for the qualitative understanding of the results in the right panels of Fig.~\ref{fig:all}. 

Quantitative results for the event-by-event averages,
$\langle \epsilon_{n} \rangle$, and the scaled standard deviations,
$\sigma(\epsilon_{n})/\langle \epsilon_{n} \rangle$, for ellipticity
and triangularity are shown in the right panels of Fig.~\ref{fig:all},
where we plot these quantities as functions of the number of the
wounded nucleons, $N_{\rm w}$.  For the average eccentricities, we can see the
advocated behavior from the orientation-multiplicity correlations in the top right panel.
We note a significant increase of
$\langle \epsilon_3 \rangle$ with $N_w$, and a corresponding decrease
of $\langle \epsilon_2 \rangle$.  For the scaled variances the
behavior is opposite, as expected from the division by
$\langle \epsilon_n \rangle$.  For the unclustered case (bottom right panel) the behavior
of $\epsilon_2$ and $\epsilon_3$ is similar, as both follow from the fluctuations only).  

The discussed orientation mechanism also leads to specific correlations of
ellipticity and triangularity for the clustered case, as displayed in Fig.~\ref{fig:corr}, where we notice a
significant anticorrelation, with the correlation coefficient
$\rho(\epsilon_2,\epsilon_3)\simeq -0.3$. The unclustered case exhibits no correlations of this kind

The final important point is the relation of the obtained shape parameters of the initial fireball to measurable 
features in the momentum distributions of produced hadrons. 
The key result here, following to the {\em collectivity} of the evolution, 
is the proportionality of the average values of the harmonic flow coefficients to 
average eccentricities,
$\langle v_n \rangle/\langle \epsilon_n \rangle = A$, with the constant $A$ increasing slowly with the 
particle multiplicity~\cite{Alver:2006wh,Abelev:2008ae,Teaney:2010vd,ALICE:2011ab,Qiu:2011iv}. Therefore, for instance, when $\langle \epsilon_3 \rangle$ increases with multiplicity, so will 
$\langle v_3 \rangle$. 
For the fluctuation measures the situation a bit more involved due to possibly large contributions to variances at the hadronization stage
from the finite number of produced hadrons. 
Nevertheless, it was found by combining experiment and model simulations that to a good approximation 
$\sigma(v_n)/\langle v_n \rangle \simeq \sigma(\epsilon_n)/\langle \epsilon_n \rangle$~\cite{Alver:2007rm,Sorensen:2008zk,Aad:2013xma}.
It means that in our case the scaled variance of the triangular flow should be significantly larger than for the elliptic flow. 

In conclusion, we list the geometric signatures of $\alpha$ clustering in ${}^{12}$C to be seen in ultra-relativistic heavy-ion collisions:
1)~Increase of $v_3$ with multiplicity.
2)~Decrease of scaled variance of $v_3$ with multiplicity.
3)~Event-by-event anticorrelation of $v_2$ and $v_3$. 
More sophisticated analysis of the predicted effects should incorporate event-by-event hydrodynamic or transport-model studies, as well as 
hadronization. Extensions to ultra-relativistic collisions of other light $\alpha$-clustered nuclei colliding on heavy nuclei 
are straightforward and will be presented elsewhere. The reason for the selection of such asymmetric collisions 
is the fact that in light nuclei the geometric deformation due to $\alpha$-clustering is large (the eccentricity parameters are 
big), while the collision with a heavy nucleus leads to fireballs which are sufficiently large and dense to exhibit well-understood collective 
behavior in the evolution, leading to harmonic flow. That is not necessarily the case for collisions of two light nuclei.

Hopefully, the possible future data in conjunction with a detailed knowledge of the dynamics of the evolution of the fireball 
will allow to place constrains on the $\alpha$-cluster structure of the 
colliding nuclei. Conversely, the knowledge of the clustered nuclear wave functions may help to test geometric patterns in fireball evolution models.  

From a broader perspective, our proposal may be viewed as an example of study of nuclear deformations/correlations via harmonic flow. 
For heavy deformed systems (U-U, Cu-Au, as measured at RHIC), certain analyses were recently proposed~\cite{Voloshin:2010ut,st}. For very light-heavy 
systems the elliptic flow has been detected in d-Au collisions~\cite{Adare:2013piz}, while the proposed studies of the triton-Au or
${}^3$He-Au collisions  at RHIC~\cite{Sickles:2013mua} should look for similar signatures as discussed in this Letter.
However, as discussed above, larger systems with intrinsic deformation, such as light nuclei with $\alpha$ clusters, create larger fireballs whose evolution 
is expected to be as collective as in the heavy-ion case, thus leading to the eccentricity--harmonic flow transmutation.

This research was supported by by the Polish National Science Centre,
grants DEC-2011/01/D/ST2/00772 and DEC-2012/06/A/ST2/00390, Spanish
DGI (grant FIS2011-24149) and Junta de Andaluc\'{\i}a (grant FQM225).
WB is grateful to Maciej Rybczy\'nski and ERA to Enrique Buend\'{\i}a for
useful conversations.

\bibliography{clusters,hydr}

\end{document}